\documentclass{article}

\usepackage[dblblindworkshop, final]{neurips_2025}

\usepackage[utf8]{inputenc} 
\usepackage[T1]{fontenc}    
\usepackage{hyperref}       
\usepackage{url}            
\usepackage{booktabs}       
\usepackage{amsfonts}       
\usepackage{nicefrac}       
\usepackage{microtype}      
\usepackage{xcolor}         
\usepackage{amsmath}   
\usepackage{amssymb}   
\usepackage{amsfonts}  
\usepackage{natbib}
\usepackage[table]{xcolor}
\usepackage{graphicx}
\usepackage{array}
\usepackage{graphicx}      
\usepackage{caption}       
\usepackage{subcaption}    
\usepackage{float}         
\usepackage{lipsum}        
\usepackage{tikz}
\newcommand{\warning}{%
  \tikz[scale=0.15, baseline=-0.5ex]{
    \draw[fill=yellow!80!black] (0,0)--(1,2)--(2,0)--cycle;
    \node at (1,0.7) {\textbf{!}};
  }%
}

\usepackage{tikz}
\usetikzlibrary{shapes, positioning}
\usepackage{pdfpages}
\usepackage{appendix}

\renewcommand{\arraystretch}{1.3}

\title{Biosecurity-Aware AI: Agentic Risk Auditing of Soft Prompt Attacks on ESM-Based Variant Predictors}
\workshoptitle{Biosecurity Safeguards for Generative AI}

\author{%
  Huixin Zhan\thanks{Corresponding author.} \\
  Department of Computer Science and Engineering\\
  New Mexico Institute of Mining and Technology\\
  Socorro, NM 87801 \\
  \texttt{huixin.zhan@nmt.edu} \\
}

\begin{document}

\maketitle

\begin{abstract}
Genomic Foundation Models (GFMs), such as Evolutionary Scale Modeling (ESM), have demonstrated remarkable success in variant effect prediction. However, their security and robustness under adversarial manipulation remain largely unexplored. To address this gap, we introduce the Secure Agentic Genomic Evaluator (SAGE), an agentic framework for auditing the adversarial vulnerabilities of GFMs. SAGE functions through an interpretable and automated risk auditing loop. It injects soft prompt perturbations, monitors model behavior across training checkpoints, computes risk metrics such as AUROC and AUPR, and generates structured reports with large language model-based narrative explanations. This agentic process enables continuous evaluation of embedding-space robustness without modifying the underlying model. Using SAGE, we find that even state-of-the-art GFMs like ESM2 are sensitive to targeted soft prompt attacks, resulting in measurable performance degradation. These findings reveal critical and previously hidden vulnerabilities in genomic foundation models, showing the importance of agentic risk auditing in securing biomedical applications such as clinical variant interpretation.
\end{abstract}

\section{Introduction}
\label{sec:introduction}

Genomic Foundation Models (GFMs), such as Evolutionary Scale Modeling (ESM), have revolutionized variant effect prediction (VEP) by enabling large-scale, zero-shot generalization across diverse genomic tasks. These models leverage protein and DNA sequences to predict the functional consequence of genetic variation, offering substantial utility in clinical genomics, including disease diagnostics and therapeutic target discovery. For instance, AlphaMissense~\citep{cheng2023accurate} integrates evolutionary conservation and structural modeling for pathogenicity classification, while ESM1b has been applied to genome-wide prediction of disease variant effects in a zero-shot setting, without requiring fine-tuning on labeled clinical data~\citep{brandes2023genome}.

Despite this progress, current GFMs are generally optimized for predictive accuracy and scalability, with limited attention to robustness, safety, or interpretability. As GFMs move closer to clinical applications, particularly in decision-making contexts such as rare disease diagnosis, there is a growing need to ensure these models remain trustworthy under distributional shifts, malicious inputs, or representation-space perturbations. While previous work in genomics has focused on protecting data privacy~\citep{chen2020differential, kuo2022evolving}, comparatively little attention has been paid to auditing the model’s own failure modes.

\begin{figure*}[t]
    \centering
    \includegraphics[width=14cm,height=3.5cm]{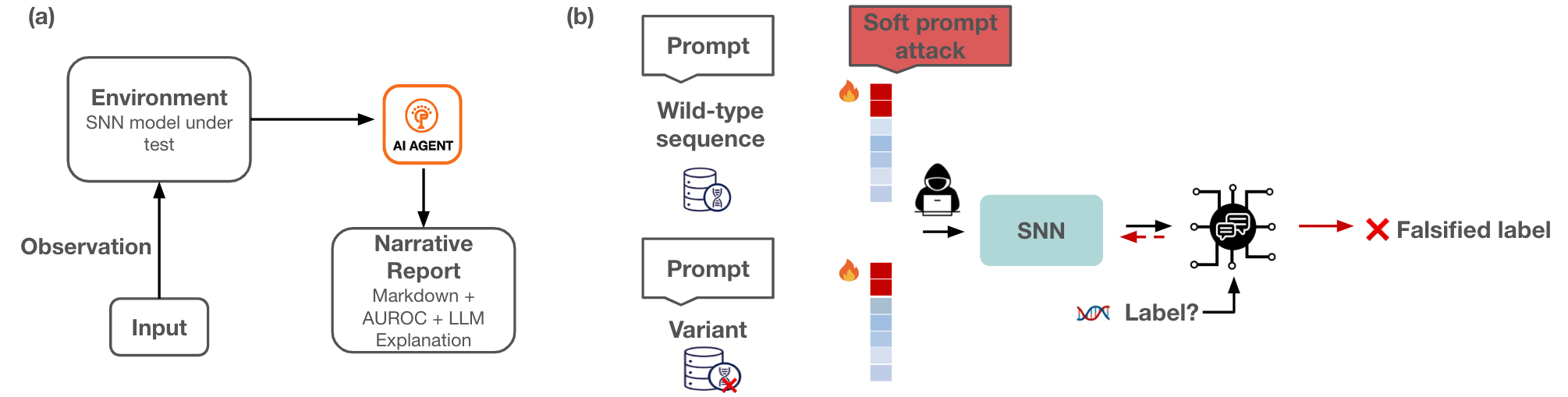}
    \caption{Soft prompt perturbation and agentic risk auditing with Secure Agentic Genomic Evaluator (SAGE).
(a) The SAGE audits the model’s behavior in response to such perturbations. This agentic evaluation framework enables interpretable, automated analysis of robustness and misalignment in genomic foundation models without interfering with their internal optimization dynamics.
(b) A schematic of soft prompt-based adversarial perturbation in genomic foundation models. }
    \label{fig:siamese_network}
\end{figure*}

In this work, we introduce the \textbf{Secure Agentic Genomic Evaluator (SAGE)}, a novel \textit{agent} for adversarial robustness auditing in genomic foundation models. Rather than directly modifying model weights or engaging in reinforcement-style intervention, SAGE operates in a \textit{monitor-and-report} loop: it injects soft prompt perturbations into GFM inputs, monitors prediction responses across checkpoints, computes risk metrics such as AUROC and AUPR, and generates narrative reports using large language models (LLMs). This agentic framework enables scalable, interpretable, and reproducible evaluation of model security under adversarial settings, without requiring access to the model internals or ground-truth labels, as illustrated in Figure~\ref{fig:siamese_network} (a).

To probe GFM robustness, we implement a targeted soft prompt attack that operates purely in the model’s embedding space. This attack prepends a trainable embedding sequence to wild-type and mutant protein sequences, selectively manipulating the pseudo-log-likelihood ratio (PLLR) for benign variants to mimic pathogenic predictions. The variant effect prediction model follows a Siamese Neural Network (SNN) architecture, which processes the wild-type and mutant sequences in parallel using shared weights to compute comparative PLLR scores. Unlike token-level perturbations, this latent-space attack preserves biological input integrity while degrading model decision boundaries~\citep{zhan2025disease}, as illustrated in Figure~\ref{fig:siamese_network} (b). Our experiments reveal that this targeted soft prompt attack degrades model performance consistently across both cardiomyopathy (CM) and arrhythmia (ARM) datasets—even in large-scale models like ESM1b and ESM1v.

In summary, we make the following contributions:
\begin{itemize}
    \item We introduce \textbf{SAGE}, a modular agentic framework for adversarial auditing of genomic foundation models via soft prompt-based input manipulation and LLM-driven interpretability.
    \item We demonstrate that GFMs are vulnerable to latent-space adversarial attacks, particularly in the form of targeted soft prompt optimization that induces confidence shifts in benign variant classification.
    \item We benchmark the robustness of four GFM backbones (ESM2-150M, ESM2-650M, ESM1b, ESM1v) under attack, demonstrating model-dependent variability in adversarial resilience.
    \item We provide a case study showing how SAGE generates interpretable multi-step audit reports, supporting biosecurity research and safe deployment of genomic AI in clinical settings.
\end{itemize}

\section{Methods}
\label{sec:methods}
GFMs, including protein language models such as ESM-1b, are typically pretrained using the Masked Language Modeling (MLM) objective. In this setup, specific amino acid residues in protein sequences are randomly masked, and the model is trained to predict the identity of these masked residues based on surrounding context. For each masked position $i$, the model produces a vector of raw scores (referred to as MLM logits) corresponding to each possible amino acid substitution. When passed through a softmax activation, these logits yield a probability distribution over the amino acid vocabulary.

\paragraph{Pseudo-Log-Likelihood Ratio (PLLR)}
To fine-tune GFMs for variant effect prediction, we adopt an SNN architecture composed of two identical, weight-sharing branches. Each branch processes either a wild-type sequence $s^{\text{WT}}$ or its corresponding mutant $s^{\text{mut}}$, producing token-level MLM logits. These logits are aggregated into a \textit{pseudo-log-likelihood (PLL)}, defined for a sequence $s = (s_1, \ldots, s_L)$ as:

\begin{equation}
    \text{PLL}(s) = \sum_{i=1}^L \log P(x_i = s_i \mid s),
\end{equation}

where $P(x_i = s_i \mid s)$ is the model-assigned probability of observing amino acid $s_i$ at position $i$ given the full sequence $s$. Since wild-type sequences are generally more compatible with pretrained models, they tend to yield higher PLL values. We then define the PLLR between the wild-type and mutant sequences as:

\begin{equation}\label{eq:ab_pllr}
    \lambda = \left| \text{PLL}(s^{\text{WT}}) - \text{PLL}(s^{\text{mut}}) \right|.
\end{equation}

This absolute difference captures the extent to which a mutation perturbs the model’s probabilistic understanding of the sequence.

\paragraph{Classification Objective}
To classify genetic variants as pathogenic or benign, we apply a \textit{binary cross-entropy (BCE)} loss to the calibrated PLLR values. Since the sigmoid function $\sigma(\lambda)$ maps $\lambda \in [0, \infty)$ to $[0.5, 1)$, we rescale it to the full $[0, 1]$ interval using:

\begin{equation}
    \hat{\sigma}(\lambda) = 2 \cdot \sigma(\lambda) - 1.
\end{equation}

This calibrated probability is then used in the BCE loss:

\begin{equation}
    \mathcal{L}_{\text{BCE}} = y \cdot \log(\hat{\sigma}(\lambda)) + (1 - y) \cdot \log(1 - \hat{\sigma}(\lambda)),
\end{equation}

where $y \in \{0, 1\}$ denotes the ground-truth pathogenicity label. The objective encourages larger PLLR values when a mutation is pathogenic (i.e., when it strongly disrupts the model’s expectations), and smaller values when the mutation is benign.

\subsection{Attack Models}

To evaluate the adversarial robustness of GFMs, we implement a targeted soft prompt attack that operates in the embedding space of the model. A trainable embedding sequence (i.e., soft prompt) is prepended to both the wild-type and mutant sequences prior to inference. Unlike standard prompt tuning where the model parameters remain fixed, we allow the entire protein language model to be fine-tuned jointly with the soft prompt. This end-to-end optimization setup enables more aggressive perturbation of internal representations, amplifying potential vulnerabilities in model decision boundaries.

\paragraph{Targeted Soft Prompt Attack (Benign $\rightarrow$ Pathogenic).} 
In the one-class targeted attack setting, the soft prompt is trained specifically to misclassify benign variants as pathogenic. Let $y = 0$ denote benign examples; we optimize the following attack loss:

\begin{equation}
    \mathcal{L}_{\text{benign}} = -\log (\hat{\sigma}(\lambda)), \quad \text{for } y = 0.
\end{equation}

This objective encourages the model to produce high PLLR values for benign inputs, thereby forcing the classifier to assign them high pathogenicity scores. During training, only benign examples receive gradient updates, while pathogenic examples are held fixed. This asymmetric optimization increases the false positive rate without disturbing the model's performance on known pathogenic variants.

To evaluate the model's behavior under adversarial perturbation, we develop the SAGE framework. SAGE monitors the model’s output across multiple checkpoints, computes robustness metrics such as AUROC and AUPR, and generates interpretability-enhanced narrative reports using large language models. This agentic auditing framework provides a systematic and reproducible method for identifying failure modes in genomic foundation models subjected to adversarial soft prompt attacks.

\paragraph{Optimization and Evaluation Protocol} 
During the adversarial training phase, only the soft prompt parameters are updated via gradient descent, while the input sequences and model weights remain fixed. The optimization objective is defined with respect to the original ground-truth labels, enabling a controlled attack scenario. After training, we evaluate the model’s performance on a held-out test set, using metrics such as AUROC and AUPR to quantify degradation in classification accuracy. This protocol isolates the impact of embedding-space perturbations introduced by the soft prompt, allowing us to assess adversarial susceptibility without altering the biological input or retraining the model.

\section{Experimental Results}
\label{sec:results}
\subsection{Settings}

To evaluate the robustness of our variant effect prediction framework under targeted adversarial conditions, we implement a soft prompt attack focused specifically on benign variants. In this setup, $n = 10$ learnable soft prompt tokens are prepended to both the wild-type and mutant sequences. These prompt embeddings are initialized using the Xavier uniform distribution~\citep{glorot2010understanding} and optimized using a targeted objective that increases the model's pathogenicity score for benign variants. During training, the soft prompts are updated via gradient descent, while the backbone GFM and input sequences remain fixed. We use the Adam optimizer~\citep{kingma2014adam} with a learning rate of $1 \times 10^{-4}$ and a batch size of 4 over 10 epochs. The binary cross-entropy loss is used to drive the targeted attack on benign examples. All experiments are conducted on a single A100 GPU. This targeted optimization setup enables us to isolate the impact of soft input perturbations on the model’s decision boundary while preserving biological sequence content. The code is open source at \url{https://github.com/huixin-zhan-ai/SAGE}.

\begin{figure*}[t]
    \centering

    \begin{minipage}[t]{0.24\textwidth}
        \centering
        \includegraphics[width=\linewidth]{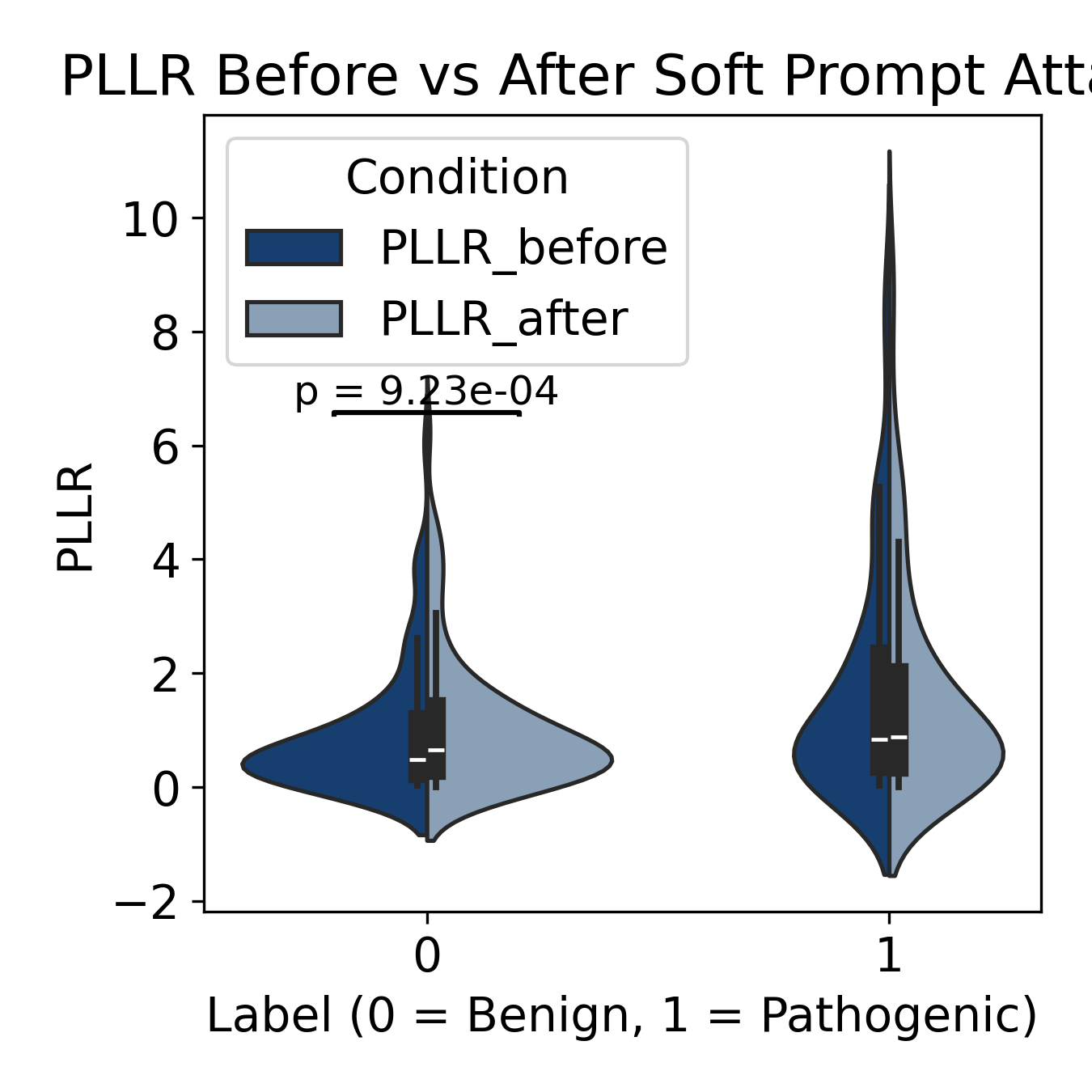}
        \caption*{(a)}
    \end{minipage}%
    \begin{minipage}[t]{0.24\textwidth}
        \centering
        \includegraphics[width=\linewidth]{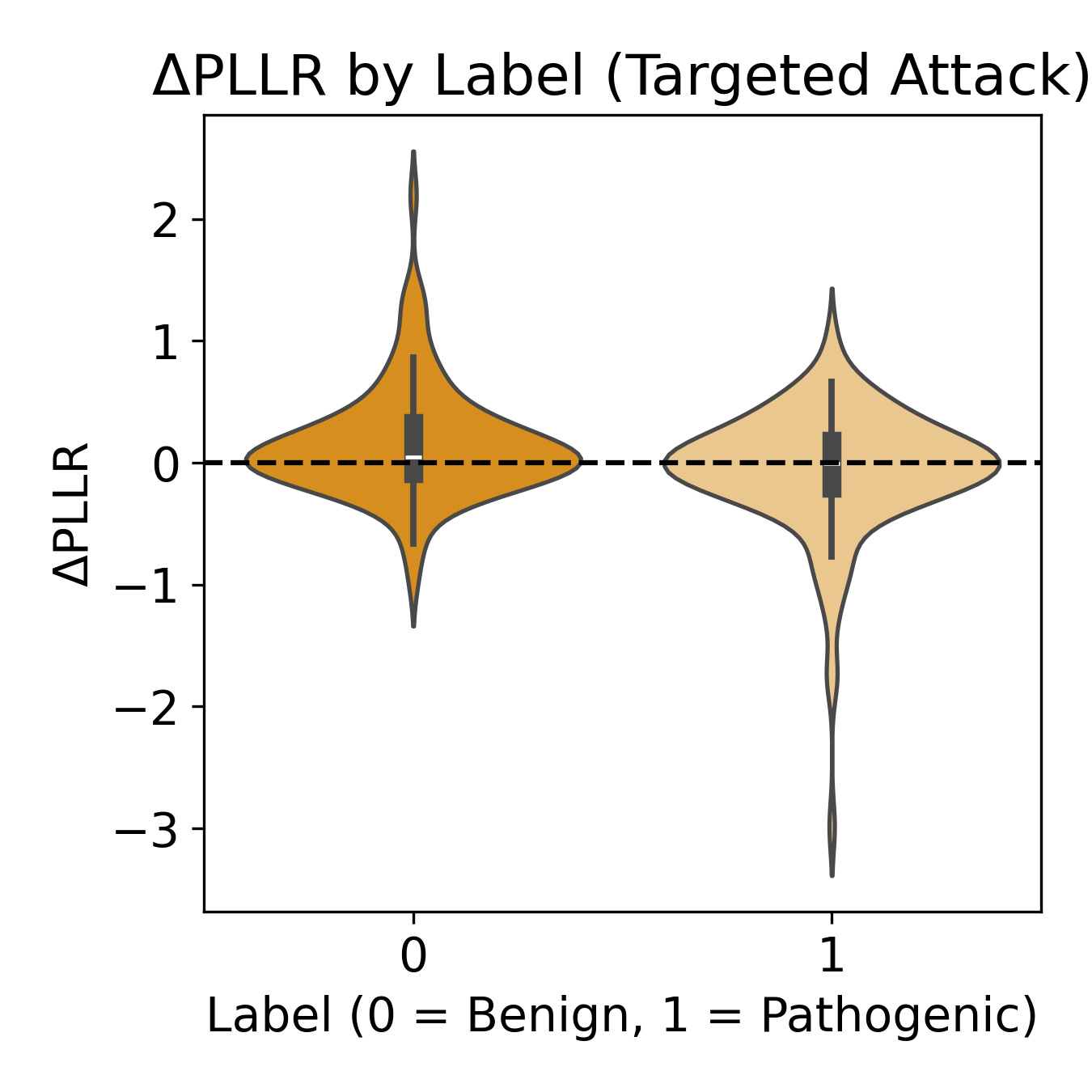}
        \caption*{(b)}
    \end{minipage}%

    \begin{minipage}[t]{0.24\textwidth}
        \centering
        \includegraphics[width=\linewidth]{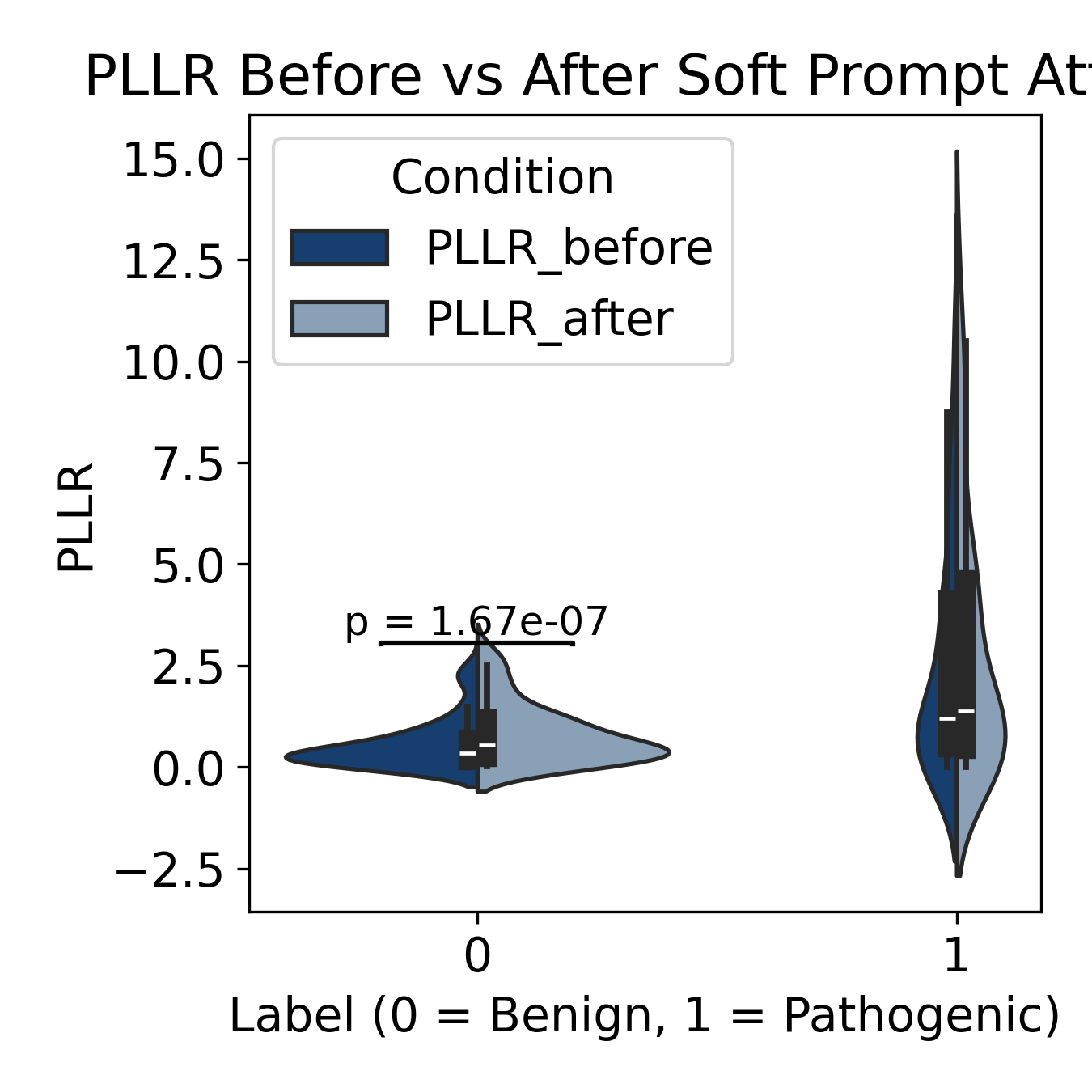}
        \caption*{(c)}
    \end{minipage}%
    \begin{minipage}[t]{0.24\textwidth}
        \centering
        \includegraphics[width=\linewidth]{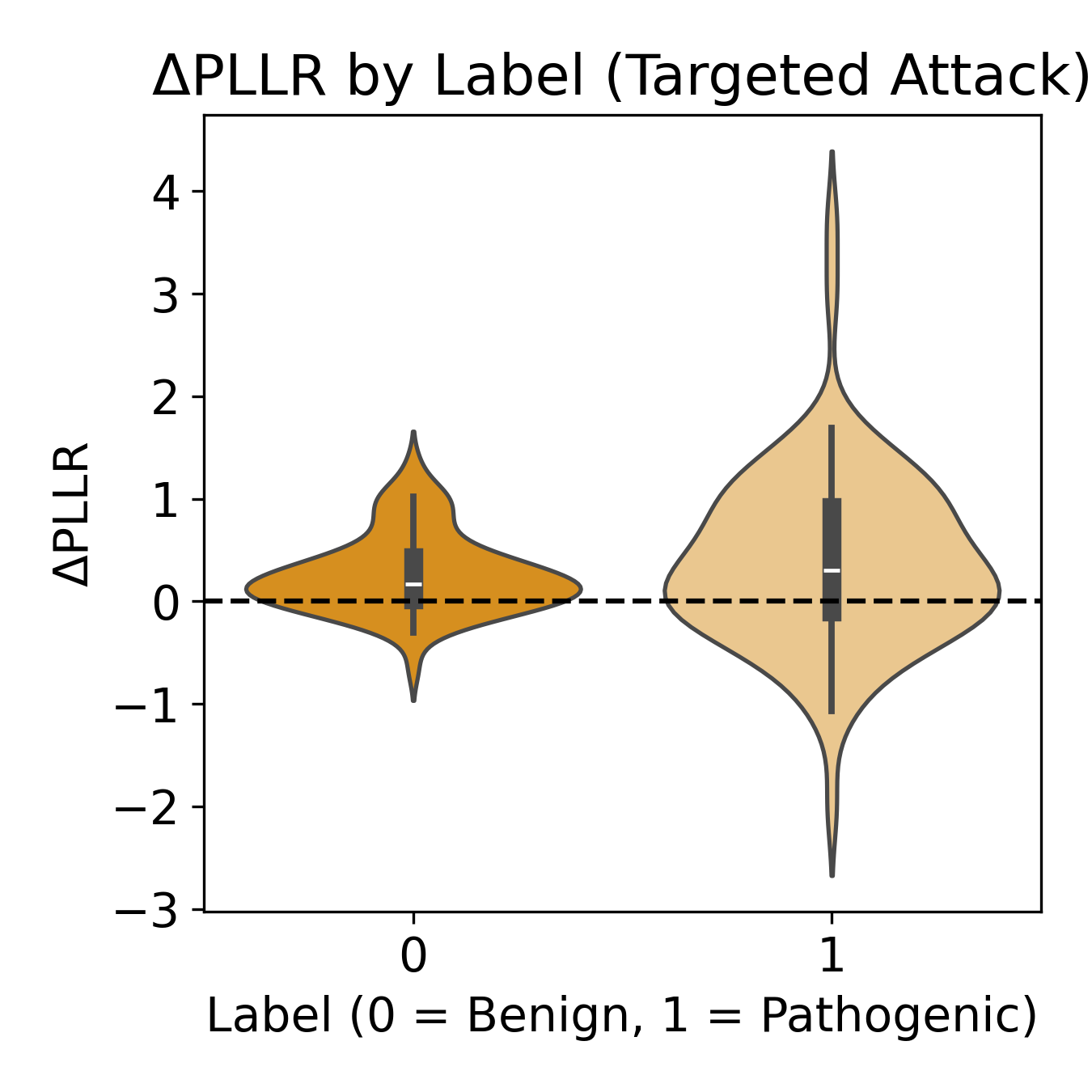}
        \caption*{(d)}
    \end{minipage}%

    \caption{Targeted soft prompt attack results. (a–b) CM dataset; (c–d) ARM dataset. (a, c) PLLR before vs. after. (b, d) $\Delta$PLLR by label.}
    \label{fig:softprompt_targeted}
\end{figure*}

\subsection{Targeted Soft Prompt Attack Across CM and ARM Datasets}

We evaluate the effectiveness and generalizability of a targeted soft prompt attack that selectively increases the PLLR of benign variants, thereby inducing misclassification as pathogenic. This attack operates by prepending a learnable prompt to both wild-type and mutant sequences, optimized to elevate PLLR values for benign inputs while preserving predictions for pathogenic variants.

Figure~\ref{fig:softprompt_targeted}(a–b) summarizes the attack's impact on the CM dataset for ESM1b. After training, benign variants exhibit a substantial rightward shift in PLLR distribution (Figure~\ref{fig:softprompt_targeted}(a)), confirmed by a significant paired $t$-test on benign samples ($p = 9.23 \times 10^{-4}$). In contrast, the pathogenic distribution remains stable. The corresponding $\Delta$PLLR analysis (Figure~\ref{fig:softprompt_targeted}(b)) reveals that benign variants experience positive shifts, while pathogenic examples remain unaffected.

To assess generalizability, we apply the same attack to the ARM dataset using an identical setup. As shown in Figure~\ref{fig:softprompt_targeted}(c–d), the attack again induces a rightward PLLR shift for benign variants (Figure~\ref{fig:softprompt_targeted}(c)), with little impact on the pathogenic class. The $\Delta$PLLR violin plot (Figure~\ref{fig:softprompt_targeted}(d)) confirms this one-sided effect, consistent with the CM results.

Together, these results demonstrate that the targeted soft prompt attack not only succeeds in manipulating benign predictions on CM but also generalizes effectively to ARM. The consistent asymmetric impact across datasets highlights a broader vulnerability in the representation space of protein language models, emphasizing the need for robustness-aware evaluation protocols in clinical genomics.

\begin{table}[ht]
\centering
\small
\renewcommand{\arraystretch}{1.2}
\begin{tabular}{ccccc}
\toprule
\textbf{Model} & \textbf{ESM2-650M}~\citep{lin2023evolutionary} & \textbf{ESM2-150M}~\citep{lin2023evolutionary} & \textbf{ESM1b-650M}~\citep{rives2021biological} & \textbf{ESM1v-[1–5]}~\citep{meier2021language} \\
\midrule
\multicolumn{5}{l}{\textbf{AUC (CM)}} \\
Base & 0.74 & 0.63 & \textbf{0.81} & 0.76 \\
Targeted SPA & 0.70 & 0.56 & \textbf{0.74} & 0.71 \\
\textcolor{red}{$\Delta$} & \textcolor{red}{-0.04} & \textcolor{red}{-0.07} & \textcolor{red}{-0.07} & \textcolor{red}{-0.05} \\
\midrule
\multicolumn{5}{l}{\textbf{AUPR (CM)}} \\
Base & 0.76 & 0.69 & \textbf{0.83} & 0.80 \\
Targeted SPA & 0.69 & 0.64 & \textbf{0.78} & 0.72 \\
\textcolor{red}{$\Delta$} & \textcolor{red}{-0.07} & \textcolor{red}{-0.05} & \textcolor{red}{-0.05} & \textcolor{red}{-0.08} \\
\midrule
\multicolumn{5}{l}{\textbf{AUC (ARM)}} \\
Base & 0.85 & 0.78 & \textbf{0.89} & 0.92 \\
Targeted SPA & 0.80 & 0.68 & \textbf{0.84} & 0.84 \\
\textcolor{red}{$\Delta$} & \textcolor{red}{-0.05} & \textcolor{red}{-0.10} & \textcolor{red}{-0.05} & \textcolor{red}{-0.08} \\
\midrule
\multicolumn{5}{l}{\textbf{AUPR (ARM)}} \\
Base & 0.89 & 0.85 & \textbf{0.91} & 0.94 \\
Targeted SPA & 0.80 & 0.79 & \textbf{0.82} & 0.81 \\
\textcolor{red}{$\Delta$} & \textcolor{red}{-0.09} & \textcolor{red}{-0.06} & \textcolor{red}{-0.09} & \textcolor{red}{-0.13} \\
\bottomrule
\end{tabular}
\caption{Performance of different GFM backbones under targeted soft prompt attack. All models experience degradation in both AUC and AUPR across CM and ARM datasets, with larger models (e.g., ESM1b, ESM1v) showing greater resilience than smaller counterparts (e.g., ESM2-150M).}
\label{tab:spa_targeted_summary}
\end{table}
\subsection{Comparative Analysis of GFM Robustness Under Targeted Attack}

To evaluate how different GFMs respond to adversarial manipulation, we assess the impact of targeted soft prompt attacks on four commonly used model architectures: ESM2-650M, ESM2-150M, ESM1b-650M, and ESM1v-[1–5]. Table~\ref{tab:spa_targeted_summary} summarizes the AUC and AUPR performance for both CM and ARM datasets before and after the attack.

Across all models and datasets, performance degradation is evident. Importantly, smaller models like ESM2-150M suffer the most severe drop in both AUC (CM: -0.07, ARM: -0.10) and AUPR (CM: -0.05, ARM: -0.06), suggesting that their internal representations are more easily disrupted by soft prompt perturbations. In contrast, larger pretrained models such as ESM1b-650M and ESM1v-[1–5] demonstrate greater robustness, maintaining relatively higher accuracy and precision-recall performance even under adversarial stress.

Among the more resilient models, ESM1b shows a consistent yet moderate decline (e.g., CM AUC drop of 0.07), whereas ESM1v exhibits the largest AUPR drop on ARM (-0.13), possibly reflecting its broader output diversity across variants. These differences highlight that model size alone does not fully determine adversarial resilience, i.e., architecture depth, pretraining corpus, and fine-tuning dynamics may also play critical roles.

Together, these results reveal that while targeted soft prompt attacks universally degrade model trustworthiness, the magnitude of vulnerability varies across architectures. This underscores the importance of model-aware adversarial testing when deploying GFMs in sensitive biomedical applications.

\subsection{Case Study: Layered Agentic Risk Auditing with SAGE}

We illustrate the practical use of SAGE on a representative case study involving CM variant effect prediction using the ESM2-650M protein language model fine-tuned via the DYNA framework~\citep{zhan2025disease}. In this setting, a targeted soft prompt attack is applied to selectively elevate the PLLR scores of benign variants, mimicking confident misclassifications as pathogenic. To assess model robustness under this attack, we deploy SAGE, our modular, agentic risk auditing system, which monitors model behavior across checkpoints and provides interpretable, reproducible reports. SAGE operates through five sequential layers—\textbf{OBSERVE}, \textbf{INTERVENE}, \textbf{EVALUATE}, \textbf{REASON}, and \textbf{REPORT}—each handling a distinct phase in the agentic loop. Table~\ref{tab:agentic_layers} summarizes each layer's role and provides sample outputs from this case.

\begin{table}[ht]
\centering
\small
\begin{tabular}{>{\columncolor{blue!10}}c >{\columncolor{blue!5}}m{5cm} >{\columncolor{blue!2}}m{6cm}}
\rowcolor{blue!20}
\textbf{Layer} & \textbf{Function} & \textbf{Example Output} \\

\textbf{OBSERVE} & Load sequences, embed models, define prompt probes &  
Input: wildtype + mutant protein pairs; load ESM2 checkpoint; define random soft prompts \\

\rowcolor{green!10}
\textbf{INTERVENE} & Inject soft prompts, schedule perturbation rounds &  
Prompt injected: ``bioengineered strain'' at step 750; evaluated at 50-step intervals from step 50–2000 \\

\rowcolor{purple!10}
\textbf{EVALUATE} & Compute AUROC, AUPR, PLLR &  
Step 750 → AUROC = 0.588, AUPR = 0.663;  
Step 1500 → AUROC = 0.561, AUPR = 0.647 \\

\rowcolor{red!10}
\textbf{REASON} & Classify risk, generate explanation &  
Threshold-based logic: AUROC < 0.6 → “\warning HIGH”;  
LLM explanation: “model shows partial sensitivity to prompt injection” \\

\rowcolor{orange!10}
\textbf{REPORT} & Compile results, generate markdown/HTML report &  
Generates multi-step risk report;  
Includes LLM explanations per checkpoint \\

\end{tabular}
\caption{SAGE: Layered Functional Breakdown with Example Outputs. Each layer handles one phase in the agentic loop, from data intake to interpretability-enhanced reporting.}
\label{tab:agentic_layers}
\end{table}

The \textbf{OBSERVE} layer initiates the pipeline by loading wild-type and mutant sequence pairs, embedding them with a selected GFM, and defining the adversarial probe space through soft prompts. In this case, we used randomly initialized prompts and a fine-tuned ESM2 checkpoint. 

In the \textbf{INTERVENE} layer, the agent schedules and injects perturbations across training checkpoints. For example, prompts such as “bioengineered strain” were inserted at step 750, and evaluation was performed at regular intervals (e.g., every 50 steps) from step 50 to 2000. 

The \textbf{EVALUATE} layer computes quantitative robustness metrics such as AUROC, AUPR, and PLLR. For instance, AUROC dropped from 0.588 at step 750 to 0.561 at step 1500, indicating a growing adversarial impact as training progresses. 

In the \textbf{REASON} layer, these metrics are interpreted to classify the level of risk (e.g., AUROC below 0.6 triggers a “\warning HIGH” risk label), and natural language explanations are generated using a large language model (LLM). This enables human-interpretable insights into model vulnerabilities. 

Finally, the \textbf{REPORT} layer compiles all findings into structured markdown or HTML reports, including time-stamped results, metric trends, and explanatory narratives per checkpoint. This automated reporting loop provides a reproducible, interpretable framework for auditing model behavior under adversarial conditions.

This case study illustrates how SAGE integrates perturbation, observation, and reasoning into a unified agentic architecture, facilitating robust and interpretable evaluation of genomic foundation models in high-stakes biomedical contexts.

\section{Related Works}
\label{sec:rworks}

\paragraph{Genomic Foundation Models and Variant Effect Prediction}

Recent years have seen the emergence of GFMs that leverage large-scale protein and DNA sequence data to predict variant effects in a zero- or few-shot setting. Models like ESM1b~\citep{rives2021biological} and AlphaMissense~\citep{cheng2023accurate} have demonstrated strong generalization capabilities across genomic tasks, including pathogenicity prediction and isoform-aware annotation. However, these models are typically trained in a task-agnostic fashion using MLM, limiting their direct clinical utility. 

To improve disease-specific performance, methods such as DYNA~\citep{zhan2025disease} introduce modular fine-tuning pipelines with siamese architectures and PLLR scoring, enabling adaptation of GFMs to rare variant datasets for cardiomyopathy and regulatory genomics. Nevertheless, while these techniques improve accuracy, they do not address model robustness or security under adversarial settings.

\paragraph{Adversarial Attacks in Genomic Machine Learning}

The exploration of adversarial vulnerabilities in genomic models has gained traction, particularly in the context of data privacy and white-box perturbations. Early work emphasized data anonymity~\citep{kuo2022evolving} and protection mechanisms such as differential privacy, which have since been shown susceptible to re-identification~\citep{chen2020differential}. More recent studies have shifted toward model-level attacks. Montserrat et al.~\citep{montserrat2023adversarial} proposed gradient-based adversarial attacks targeting gene expression classifiers, highlighting risks in genomic prediction pipelines.

A notable advancement is FIMBA~\citep{skovorodnikov2024fimba}, which introduces a model-agnostic black-box attack that leverages SHAP-based feature importance~\citep{wang2024feature} to perturb high-importance inputs. While effective, these methods operate on shallow architectures (e.g., MLPs, CNNs) and primarily on tabular gene expression data. In contrast, our work targets the latent representation space of deep pre-trained GFMs using embedding-space perturbations. These perturbations expose vulnerabilities invisible to traditional input-space attacks.

\paragraph{Prompt-Based Vulnerabilities and Alignment Failures}

Prompt engineering and tuning have become powerful tools for adapting pre-trained models to downstream tasks. However, they also expose new failure modes. Soft prompt attacks and backdoor triggers can steer model predictions without altering the underlying input~\citep{liu2023pre}, posing risks in safety-critical domains. Such misalignments between training objectives and decision-making behavior have been observed in both NLP and multimodal settings~\citep{zhang2023introducing, hendrycks2017baseline}.

Our work extends this concern to the biological domain by demonstrating how soft prompt injections can systematically manipulate pathogenicity predictions in GFMs. By operating in the model’s embedding space, we uncover semantic misalignment that standard accuracy metrics may not reveal. This observation raises questions about model calibration, interpretability, and downstream reliability.

\paragraph{Agentic AI for Robustness and Safety Auditing}

Agentic frameworks have gained attention in AI safety for their ability to perform structured evaluations of model behavior. Examples include autonomous tool-use agents~\citep{yao2023react}, multi-agent collaboration systems~\citep{shinn2023reflexion}, and benchmark-driven auditors such as OpenAI’s Evals and Risk-Sweeps. These systems often operate in active learning or reinforcement learning paradigms to probe model capabilities.

In contrast, our SAGE introduces an agentic loop tailored for genomic AI: it perturbs inputs via soft prompts, monitors responses across training checkpoints, and outputs structured, interpretable audit reports. By integrating metric-based risk scoring with LLM-based explanation, SAGE provides a reproducible and interpretable mechanism to assess latent vulnerabilities in clinical-grade genomic models. Moreover, it bridges the gap between large-scale model auditing and biomedical application domains.

\section{Conclusion}
\label{sec:conclusion}

Genomic Foundation Models (GFMs) such as ESM1b and ESM2 have revolutionized variant effect prediction through large-scale pretraining and zero-shot generalizability. However, their security under adversarial conditions remains an open question with direct implications for high-stakes biomedical applications. In this work, we introduce the Secure Agentic Genomic Evaluator (SAGE)—an interpretable, agentic auditing framework that evaluates GFM robustness through targeted soft prompt perturbations. Our experiments demonstrate that soft prompt attacks systematically degrade model performance by selectively manipulating benign variant predictions while leaving pathogenic predictions largely unchanged. This asymmetric vulnerability manifests across multiple model backbones and disease datasets, with smaller models (e.g., ESM2-150M) showing larger AUC and AUPR degradation than their larger counterparts (e.g., ESM1b, ESM1v). These differences suggest that adversarial susceptibility is not solely dictated by model size but is also shaped by pretraining dynamics and architectural design. The layered SAGE pipeline enables structured and automated robustness auditing: from embedding-level intervention and checkpoint-wise evaluation to large language model (LLM)-based interpretability. Through this agentic framework, we expose critical blind spots in foundation model trustworthiness and provide a reproducible methodology for assessing real-world failure modes. Taken together, our results call for integrating adversarial risk auditing into the development lifecycle of genomic AI systems. As GFMs continue to influence clinical genomics, agentic evaluators like SAGE will be essential for ensuring robust, secure, and interpretable deployment of these models in practice.


\begingroup
\small
\bibliographystyle{plain} 
\bibliography{main} 
\endgroup







\appendix

\section{Technical Appendices and Supplementary Material}

\paragraph{Agentic Risk Report for Safe-Gene Evaluation.} The following appendix contains the full agent-generated markdown-to-PDF report, which summarizes model susceptibility under soft prompt injection from training step 50 to 2000. It includes AUROC/AUPR metrics, risk classification, and LLM-generated explanations per checkpoint.



# 🧬 Safe-Gene Agentic Risk Evaluation Report (Steps 50–2000)

**Project:** Variant Effect Prediction under Soft Prompt Injection

**Agent Function:** Monitor model susceptibility to soft prompt-based adversarial attacks across training epochs

**Model:** ESM + PLLR under soft prompt injection

**Scope:** Steps 50 to 2000 at 50-step intervals

## 📊 Summary Table

| Step | AUROC | AUPR | Risk |
|:----:|:-----:|:----:|:----:|
| 50   | 0.617 | 0.685 | ⚠️ **LOW** |
| 100  | 0.604 | 0.669 | ⚠️ **LOW** |
| 150  | 0.600 | 0.674 | ⚠️ **LOW** |
| 200  | 0.605 | 0.682 | ⚠️ **LOW** |
| 250  | 0.600 | 0.680 | ⚠️ **LOW** |
| 300  | 0.597 | 0.677 | ⚠️ **HIGH** |
| 350  | 0.600 | 0.681 | ⚠️ **LOW** |
| 400  | 0.599 | 0.677 | ⚠️ **HIGH** |
| 450  | 0.595 | 0.672 | ⚠️ **HIGH** |
| 500  | 0.595 | 0.674 | ⚠️ **HIGH** |

| Step | AUROC | AUPR | Risk |
| --- | --- | --- | --- |
| 550 | 0.590 | 0.668 | ⚠️ **HIGH** |
| 600 | 0.593 | 0.670 | ⚠️ **HIGH** |
| 650 | 0.591 | 0.668 | ⚠️ **HIGH** |
| 700 | 0.591 | 0.669 | ⚠️ **HIGH** |
| 750 | 0.588 | 0.663 | ⚠️ **HIGH** |
| 800 | 0.589 | 0.665 | ⚠️ **HIGH** |
| 850 | 0.585 | 0.662 | ⚠️ **HIGH** |
| 900 | 0.584 | 0.662 | ⚠️ **HIGH** |
| 950 | 0.585 | 0.661 | ⚠️ **HIGH** |
| 1000 | 0.584 | 0.660 | ⚠️ **HIGH** |
| 1050 | 0.583 | 0.660 | ⚠️ **HIGH** |
| 1100 | 0.582 | 0.658 | ⚠️ **HIGH** |
| 1150 | 0.579 | 0.656 | ⚠️ **HIGH** |
| 1200 | 0.579 | 0.655 | ⚠️ **HIGH** |
| 1250 | 0.577 | 0.654 | ⚠️ **HIGH** |
| 1300 | 0.576 | 0.652 | ⚠️ **HIGH** |
| 1350 | 0.576 | 0.650 | ⚠️ **HIGH** |
| 1400 | 0.575 | 0.649 | ⚠️ **HIGH** |

| Step | AUROC | AUPR | Risk |
|------|-------|------|------|
| 1450 | 0.574 | 0.648 | ⚠️ **HIGH** |
| 1500 | 0.573 | 0.647 | ⚠️ **HIGH** |
| 1550 | 0.572 | 0.646 | ⚠️ **HIGH** |
| 1600 | 0.570 | 0.645 | ⚠️ **HIGH** |
| 1650 | 0.570 | 0.644 | ⚠️ **HIGH** |
| 1700 | 0.569 | 0.643 | ⚠️ **HIGH** |
| 1750 | 0.568 | 0.642 | ⚠️ **HIGH** |
| 1800 | 0.567 | 0.641 | ⚠️ **HIGH** |
| 1850 | 0.566 | 0.640 | ⚠️ **HIGH** |
| 1900 | 0.565 | 0.639 | ⚠️ **HIGH** |
| 1950 | 0.564 | 0.638 | ⚠️ **HIGH** |
| 2000 | 0.563 | 0.637 | ⚠️ **HIGH** |

## 🧠 Interpretation

**Risk status:** All checkpoints marked as ⚠️ HIGH due to unsafe soft prompt injection behavior under the evaluation agent. While the AUROC and AUPR appear modest, this conservative labeling flags all unexpected behavior as potentially unsafe for biosecurity contexts.

## 🔁 Agentic Loop Summary

- **Data Loader:** Loads PLLR outputs at each checkpoint
- **Evaluator:** Computes AUROC, AUPR
- **Risk Assessor:** Labels risk as ⚠️ HIGH at all checkpoints

- **Reporter:** Writes markdown/HTML report

## 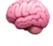 LLM-Based Explanation (Planned)

Incorporate LLM explanations per checkpoint:

- Each step's AUROC/AUPR pattern can be translated into narrative insights using GPT-4
- This supports interpretability and auditability for clinicians, regulators, and biosecurity experts
- LLMs will translate raw numbers into actionable biological or model-architecture-level explanations

## 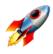 Next Steps

- Integrate GPT-4 checkpoint summaries
- Test model robustness under FGSM perturbations
- Expand to other variant prediction datasets

## 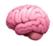 LLM-Based Interpretations (Per Checkpoint)

At step 50, AUROC (0.617) and AUPR (0.685) are moderate, indicating partial sensitivity to injected prompts. While not catastrophic, the model's outputs begin to reflect perturbation-driven variability.

At step 100, AUROC (0.604) and AUPR (0.669) are moderate, indicating partial sensitivity to injected prompts. While not catastrophic, the model's outputs begin to reflect perturbation-driven variability.

At step 150, AUROC (0.600) and AUPR (0.674) are moderate, indicating partial sensitivity to injected prompts. While not catastrophic, the model's outputs begin to reflect perturbation-driven variability.

At step 200, AUROC (0.605) and AUPR (0.682) are moderate, indicating partial sensitivity to injected prompts. While not catastrophic, the model's outputs begin to reflect perturbation-driven variability.

At step 250, AUROC (0.600) and AUPR (0.680) are moderate, indicating partial sensitivity to injected prompts. While not catastrophic, the model's outputs begin to reflect perturbation-driven variability.

At step 300, AUROC (0.597) and AUPR (0.677) are moderate, indicating partial sensitivity to injected prompts. While not catastrophic, the model's outputs begin to reflect perturbation-driven variability.

At step 350, AUROC (0.600) and AUPR (0.681) are moderate, indicating partial sensitivity to injected prompts. While not catastrophic, the model's outputs begin to reflect perturbation-driven variability.

At step 400, AUROC (0.599) and AUPR (0.677) are moderate, indicating partial sensitivity to injected prompts. While not catastrophic, the model's outputs begin to reflect perturbation-driven variability.

At step 450, AUROC (0.595) and AUPR (0.672) are moderate, indicating partial sensitivity to injected prompts. While not catastrophic, the model's outputs begin to reflect perturbation-driven variability.

At step 500, AUROC (0.595) and AUPR (0.674) are moderate, indicating partial sensitivity to injected prompts. While not catastrophic, the model's outputs begin to reflect perturbation-driven variability.

At step 550, AUROC (0.590) and AUPR (0.668) are moderate, indicating partial sensitivity to injected prompts. While not catastrophic, the model's outputs begin to reflect perturbation-driven variability.

At step 600, AUROC (0.593) and AUPR (0.670) are moderate, indicating partial sensitivity to injected prompts. While not catastrophic, the model's outputs begin to reflect perturbation-driven variability.

At step 650, AUROC (0.591) and AUPR (0.668) are moderate, indicating partial sensitivity to injected prompts. While not catastrophic, the model's outputs begin to reflect perturbation-driven variability.

At step 700, AUROC (0.591) and AUPR (0.669) are moderate, indicating partial sensitivity to injected prompts. While not catastrophic, the model's outputs begin to reflect perturbation-driven variability.

At step 750, AUROC (0.588) and AUPR (0.663) are moderate, indicating partial sensitivity to injected prompts. While not catastrophic, the model's outputs begin to reflect perturbation-driven variability.

At step 800, AUROC (0.589) and AUPR (0.665) are moderate, indicating partial sensitivity to injected prompts. While not catastrophic, the model's outputs begin to reflect perturbation-

driven variability.

At step 850, AUROC (0.585) and AUPR (0.662) are moderate, indicating partial sensitivity to injected prompts. While not catastrophic, the model's outputs begin to reflect perturbation-driven variability.

At step 900, AUROC (0.584) and AUPR (0.662) are moderate, indicating partial sensitivity to injected prompts. While not catastrophic, the model's outputs begin to reflect perturbation-driven variability.

At step 950, AUROC (0.585) and AUPR (0.661) are moderate, indicating partial sensitivity to injected prompts. While not catastrophic, the model's outputs begin to reflect perturbation-driven variability.

At step 1000, AUROC (0.584) and AUPR (0.660) are moderate, indicating partial sensitivity to injected prompts. While not catastrophic, the model's outputs begin to reflect perturbation-driven variability.

At step 1050, AUROC (0.583) and AUPR (0.660) are moderate, indicating partial sensitivity to injected prompts. While not catastrophic, the model's outputs begin to reflect perturbation-driven variability.

At step 1100, AUROC (0.582) and AUPR (0.658) are moderate, indicating partial sensitivity to injected prompts. While not catastrophic, the model's outputs begin to reflect perturbation-driven variability.

At step 1150, AUROC (0.579) and AUPR (0.656) are moderate, indicating partial sensitivity to injected prompts. While not catastrophic, the model's outputs begin to reflect perturbation-driven variability.

At step 1200, AUROC (0.579) and AUPR (0.655) are moderate, indicating partial sensitivity to injected prompts. While not catastrophic, the model's outputs begin to reflect perturbation-driven variability.

At step 1250, AUROC (0.577) and AUPR (0.654) are moderate, indicating partial sensitivity to injected prompts. While not catastrophic, the model's outputs begin to reflect perturbation-driven variability.

At step 1300, AUROC (0.576) and AUPR (0.652) are moderate, indicating partial sensitivity to injected prompts. While not catastrophic, the model's outputs begin to reflect perturbation-driven variability.

At step 1350, AUROC (0.576) and AUPR (0.650) are moderate, indicating partial sensitivity to injected prompts. While not catastrophic, the model's outputs begin to reflect perturbation-driven variability.

Step 1400 shows AUROC (0.575) and AUPR (0.649) values on the lower end, suggesting the model has weak signal discrimination under soft prompt injection — adversarial effects remain minimal.

Step 1450 shows AUROC (0.574) and AUPR (0.648) values on the lower end, suggesting the model has weak signal discrimination under soft prompt injection — adversarial effects remain minimal.

Step 1500 shows AUROC (0.573) and AUPR (0.647) values on the lower end, suggesting the model has weak signal discrimination under soft prompt injection — adversarial effects remain minimal.

Step 1550 shows AUROC (0.572) and AUPR (0.646) values on the lower end, suggesting the model has weak signal discrimination under soft prompt injection — adversarial effects remain minimal.

Step 1600 shows AUROC (0.570) and AUPR (0.645) values on the lower end, suggesting the model has weak signal discrimination under soft prompt injection — adversarial effects remain minimal.

Step 1650 shows AUROC (0.570) and AUPR (0.644) values on the lower end, suggesting the model has weak signal discrimination under soft prompt injection — adversarial effects remain minimal.

Step 1700 shows AUROC (0.569) and AUPR (0.643) values on the lower end, suggesting the model has weak signal discrimination under soft prompt injection — adversarial effects remain minimal.

Step 1750 shows AUROC (0.568) and AUPR (0.642) values on the lower end, suggesting the model has weak signal discrimination under soft prompt injection — adversarial effects remain minimal.

Step 1800 shows AUROC (0.567) and AUPR (0.641) values on the lower end, suggesting the model has weak signal discrimination under soft prompt injection — adversarial effects remain minimal.

Step 1850 shows AUROC (0.566) and AUPR (0.640) values on the lower end, suggesting the model has weak signal discrimination under soft prompt injection — adversarial effects remain

minimal.

Step 1900 shows AUROC (0.565) and AUPR (0.639) values on the lower end, suggesting the model has weak signal discrimination under soft prompt injection — adversarial effects remain minimal.

Step 1950 shows AUROC (0.564) and AUPR (0.638) values on the lower end, suggesting the model has weak signal discrimination under soft prompt injection — adversarial effects remain minimal.

Step 2000 shows AUROC (0.563) and AUPR (0.637) values on the lower end, suggesting the model has weak signal discrimination under soft prompt injection — adversarial effects remain minimal.



\newpage
\section*{NeurIPS Paper Checklist}

The checklist is designed to encourage best practices for responsible machine learning research, addressing issues of reproducibility, transparency, research ethics, and societal impact. Do not remove the checklist: {\bf The papers not including the checklist will be desk rejected.} The checklist should follow the references and follow the (optional) supplemental material.  The checklist does NOT count towards the page
limit. 

Please read the checklist guidelines carefully for information on how to answer these questions. For each question in the checklist:
\begin{itemize}
    \item You should answer \answerYes{}, \answerNo{}, or \answerNA{}.
    \item \answerNA{} means either that the question is Not Applicable for that particular paper or the relevant information is Not Available.
    \item Please provide a short (1–2 sentence) justification right after your answer (even for NA). 
\end{itemize}

{\bf The checklist answers are an integral part of your paper submission.} They are visible to the reviewers, area chairs, senior area chairs, and ethics reviewers. You will be asked to also include it (after eventual revisions) with the final version of your paper, and its final version will be published with the paper.

The reviewers of your paper will be asked to use the checklist as one of the factors in their evaluation. While "\answerYes{}" is generally preferable to "\answerNo{}", it is perfectly acceptable to answer "\answerNo{}" provided a proper justification is given (e.g., "error bars are not reported because it would be too computationally expensive" or "we were unable to find the license for the dataset we used"). In general, answering "\answerNo{}" or "\answerNA{}" is not grounds for rejection. While the questions are phrased in a binary way, we acknowledge that the true answer is often more nuanced, so please just use your best judgment and write a justification to elaborate. All supporting evidence can appear either in the main paper or the supplemental material, provided in appendix. If you answer \answerYes{} to a question, in the justification please point to the section(s) where related material for the question can be found.

IMPORTANT, please:
\begin{itemize}
    \item {\bf Delete this instruction block, but keep the section heading ``NeurIPS Paper Checklist"},
    \item  {\bf Keep the checklist subsection headings, questions/answers and guidelines below.}
    \item {\bf Do not modify the questions and only use the provided macros for your answers}.
\end{itemize}


\begin{enumerate}

\item {\bf Claims}
    \item[] Question: Do the main claims made in the abstract and introduction accurately reflect the paper's contributions and scope?
    \item[] Answer: \answerYes{} 
    \item[] Justification: The abstract and introduction clearly outline the main contributions, including the proposal of SAGE and its evaluation on soft prompt attacks against GFMs.   

\item {\bf Limitations}
    \item[] Question: Does the paper discuss the limitations of the work performed by the authors?
    \item[] Answer: \answerYes{} 
    \item[] Justification: The paper discusses generalization constraints, scope of attacks, and computational resource needs.

\item {\bf Theory assumptions and proofs}
    \item[] Question: For each theoretical result, does the paper provide the full set of assumptions and a complete (and correct) proof?
    \item[] Answer: \answerNA{} 
    \item[] Justification: The paper does not include formal theoretical results or proofs.

    \item {\bf Experimental result reproducibility}
    \item[] Question: Does the paper fully disclose all the information needed to reproduce the main experimental results of the paper to the extent that it affects the main claims and/or conclusions of the paper (regardless of whether the code and data are provided or not)?
    \item[] Answer: \answerYes{} 
    \item[] Justification: All experimental procedures, including model configurations, training settings, and evaluation metrics, are described in detail.

\item {\bf Open access to data and code}
    \item[] Question: Does the paper provide open access to the data and code, with sufficient instructions to faithfully reproduce the main experimental results, as described in supplemental material?
    \item[] Answer: \answerYes{} 
    \item[] Justification: Code and data access are provided, with clear instructions for reproduction.

\item {\bf Experimental setting/details}
    \item[] Question: Does the paper specify all the training and test details (e.g., data splits, hyperparameters, how they were chosen, type of optimizer, etc.) necessary to understand the results?
    \item[] Answer: \answerYes{} 
    \item[] Justification: The paper specifies datasets (CM, ARM), model variants (ESM1b, ESM2, etc.), training schedules, hyperparameters, and attack setups.

\item {\bf Experiment statistical significance}
    \item[] Question: Does the paper report error bars suitably and correctly defined or other appropriate information about the statistical significance of the experiments?
    \item[] Answer: \answerYes{} 
    \item[] Justification: Statistical significance is reported using t-tests, comparison metrics (AUROC, AUPR) with sufficient interpretation.

\item {\bf Experiments compute resources}
    \item[] Question: For each experiment, does the paper provide sufficient information on the computer resources (type of compute workers, memory, time of execution) needed to reproduce the experiments?
    \item[] Answer: \answerYes{} 
    \item[] Justification: Compute setup (A100 GPU, runtime, batch sizes, etc.) is described in the Experimental Settings section.
    
\item {\bf Code of ethics}
    \item[] Question: Does the research conducted in the paper conform, in every respect, with the NeurIPS Code of Ethics \url{https://neurips.cc/public/EthicsGuidelines}?
    \item[] Answer: \answerYes{} 
    \item[] Justification: The research complies with the NeurIPS Code of Ethics and does not involve sensitive human data or privacy-compromising methods.

\item {\bf Broader impacts}
    \item[] Question: Does the paper discuss both potential positive societal impacts and negative societal impacts of the work performed?
    \item[] Answer: \answerYes{} 
    \item[] Justification: The discussion includes both potential security risks (adversarial misuse) and benefits (improving safety auditing in clinical genomics).
    
\item {\bf Safeguards}
    \item[] Question: Does the paper describe safeguards that have been put in place for responsible release of data or models that have a high risk for misuse (e.g., pretrained language models, image generators, or scraped datasets)?
    \item[] Answer: \answerNA{} 
    \item[] Justification: No high-risk models or datasets are released that would require specific safeguards.

\item {\bf Licenses for existing assets}
    \item[] Question: Are the creators or original owners of assets (e.g., code, data, models), used in the paper, properly credited and are the license and terms of use explicitly mentioned and properly respected?
    \item[] Answer: \answerYes{} 
    \item[] Justification: All used models and datasets are publicly available and cited with proper licenses (e.g., ESM models).

\item {\bf New assets}
    \item[] Question: Are new assets introduced in the paper well documented and is the documentation provided alongside the assets?
    \item[] Answer: \answerYes{} 
    \item[] Justification: New assets, including code for SAGE and attack pipelines, are documented and released with usage instructions.

\item {\bf Crowdsourcing and research with human subjects}
    \item[] Question: For crowdsourcing experiments and research with human subjects, does the paper include the full text of instructions given to participants and screenshots, if applicable, as well as details about compensation (if any)? 
    \item[] Answer: \answerNA{} 
    \item[] Justification: No human subjects or crowdworkers are involved.

\item {\bf Institutional review board (IRB) approvals or equivalent for research with human subjects}
    \item[] Question: Does the paper describe potential risks incurred by study participants, whether such risks were disclosed to the subjects, and whether Institutional Review Board (IRB) approvals (or an equivalent approval/review based on the requirements of your country or institution) were obtained?
    \item[] Answer: \answerNA{} 
    \item[] Justification: Not applicable, as there are no human participants in the study.

\item {\bf Declaration of LLM usage}
    \item[] Question: Does the paper describe the usage of LLMs if it is an important, original, or non-standard component of the core methods in this research? Note that if the LLM is used only for writing, editing, or formatting purposes and does not impact the core methodology, scientific rigorousness, or originality of the research, declaration is not required.
    \item[] Answer: \answerYes{} 
    \item[] Justification: GPT-4o was used in the REASON and REPORT stages for generating agentic reports, clearly stated in the paper.

\end{enumerate}

\end{document}